\documentclass[12pt]{spieman}  
\usepackage{amsmath,amsfonts,amssymb}
\usepackage{graphicx}
\usepackage{setspace}
\usepackage{tocloft}
\usepackage{soul}

\title{A Multi-station Meteor Monitoring (M$^3$) System. I. Design and Testing}

\author[a,b]{Zhenye Li}
\author[a,*]{Hu Zou}
\author[c,a,d,*]{Jifeng Liu}
\author[a,b]{Jun Ma}
\author[a,b]{Xinlin Zhao}
\author[a,b]{Xue Li}
\author[a,b]{Zhijun Tu}
\author[a,b]{Bowen Zhang}
\author[a]{Rui Wang}
\author[e]{Shaohan Wang}
\author[f]{Marco Xue}

\affil[a]{National Astronomical Observatories, Chinese Academy of Sciences, CAS Key Laboratory of Optical Astronomy,  Beijing, P.R. China, 100101}
\affil[b]{University of Chinese Academy of Sciences, School of Astronomy and Space Science, Beijing, P.R. China, 100049}
\affil[c]{National Astronomical Observatories, Chinese Academy of Sciences, New Cornerstone Science Laboratory,  Beijing, P.R. China, 100101}
\affil[d]{Beijing Normal University, Institute for Frontiers in Astronomy and Astrophysics, Beijing, P.R. China, 102206}
\affil[e]{Department of Astronomy, University of Science and Technology of China, Hefei, P.R. China, 230026}
\affil[f]{Tsinghua International School, Beijing, P.R. China, 100084}

\cftpagenumbersoff{figure}
\cftpagenumbersoff{table} 
\begin{document} 
\maketitle

\begin{abstract}
Meteors carry important and indispensable information about the interplanetary environment, which can be used to understand the origin and evolution of our solar system. We have developed a Multi-station Meteor Monitoring ($\rm M^3$) system that can observe almost the entire sky and detect meteors automatically, and it determines their trajectories. They are highly extensible to construct a large-scale network. Each station consists of a waterproof casing, a wide field-of-view lens with a CMOS camera, and a supporting computer. The camera has a built-in GPS module for accurately timing the meteoroid entry into the atmosphere (accurate to 1 {\textmu}s), which is the most prominent characteristic compared with other existing meteor monitoring devices. We have also developed a software package that can efficiently identify and measure meteors appearing in the real-time video stream and compute the orbits of meteoroids in the solar system via multi-station observations. During the Geminid meteor shower in 2021, the M$^3$ system was tested at two stations ($\sim$55 km apart) in the suburbs of Beijing. The test results show that the astrometric accuracy is about 0.3-0.4 arcmin. About 800 meteors were detected by these two stations. A total of 473 meteors have their orbits calculated by our software, and 377 of them belong to the Geminid meteoroid stream. Our M$^3$ system will be further tested and upgraded, and it will be used to construct a large monitoring network in China in the future.
\end{abstract}


{\noindent \footnotesize\textbf{*}Hu Zou,  \linkable{zouhu@nao.cas.cn} Jifeng liu,  \linkable{jfliu@nao.cas.cn}}

\begin{spacing}{2}   

\section{Introduction}
\label{sect:intro}  
Meteors are atmospheric phenomena resulting from small particles in the solar system colliding with the the Earth's atmosphere. It glows due to ionization of atoms at altitudes roughly between 30-100 km \cite{1998SSRv...84..327C}. Such particles are known as meteoroids, traveling at 11-72 ${\rm kms^{-1}}$ relative to the Earth \cite{1998SSRv...84..327C}. The meteoroids originate mainly from asteroids and comets orbiting the Sun. Asteroids may collide with each other and form smaller particles, while comets are heated by the Sun, ejecting particles in the process. These particles accumulate along their orbits into meteoroid streams, which are responsible for meteor showers. 

Meteoroids are important for understanding the activity and evolution of small objects in our solar system. It was found that 85\% of the particles forming the zodiacal light originate from Jupiter-family comets, and they can be also identified as micrometeoroids which might impact artificial satellites \cite{2010ApJ...713..816N}. Meteor trajectories determined from multi-station video observations can be used to search and identify new meteoroid streams, and subsequently find their parent bodies \cite{2011Icar..216...40J}. In addition, meteor outbursts can trace long-period comets (LPC) that were never recorded, and even help to identify Earth-threatening comets \cite{2018pimo.conf...65D}. Hazardous LPCs may only be discovered 6-12 months before their potential impacts, so continuously monitoring such meteor outbursts is helpful for planetary defense \cite{2003Icar..162..443L}.

Large meteoroids can be detected by telescopes with moderate apertures when they approach the Earth. 2022 WJ1, 2023 CX1, and 2024 BX1 are some of the asteroids that are detected by telescopes prior to the impacts with the Earth\cite{2023DPS....5551002K,2023LPICo2851.2049V,2024JIMO...52...29R}. However, smaller meteoroids can only be discovered during their entry into the Earth's atmosphere. When a meteoroid is large and slow enough to survive its entry, the remnant is called as a meteorite. However, meteorites with known orbits are still rare. As of February 20, 2023, only 46 meteorites with photographic orbits are published (\url{https://www.meteoriteorbits.info/}). 

The meteors are mainly observed in radio and optical wavelengths \cite{1998SSRv...84..327C}. The optical observation mode includes still frame and video monitoring. In decades past, still frame cameras and rotary shutters have been widely used and the temporal information of meteors were obtained by the shutter alternately blocking the light several times per second \cite{2002ESASP.500..257S}. However, with the advancement of CMOS cameras, video observations became prevalent. The CAMS network \cite{2011Icar..216...40J} and Japanese Sonotaco network \cite{2010epsc.conf..798S} have long history on detecting meteors using video cameras, date back to 2000s. The Global Meteor Network (GMN) is one of the largest meteor networks \cite{2021MNRAS.506.5046V}, whose origins trace back to the Croatian Meteor Network \cite{2018JIMO...46...87S} (\url{https://github.com/CroatianMeteorNetwork}). It is equipped with Raspberry Pi computers and inexpensive CMOS cameras, and an open source software (RMS \cite{2021MNRAS.506.5046V}, \url{ https://github.com/CroatianMeteorNetwork/RMS}). The Fireball Recovery and Inter Planetary Observation Network (FRIPON) is an active network mainly located in Europe which includes 105 optical and 25 radio stations \cite{2020A&A...644A..53C}. It also uses sets of low-cost computers and cameras, and each camera covers the full sky using a fish-eye lens. An associated detecting software, \textit{FreeTure} (\url{https://github.com/fripon/freeture}), was specially developed for this network. In addition, the commercial software developed by Sonotaco, \textit{UFOCapture} (\url{http://sonotaco.com/soft/UFO2/help/english/index.html}), is widely used among smaller networks\cite{2008epsc.conf..738M,2014EGUGA..16.4906Y,2013JIMO...41...84S,2008EM&P..102..263A} and amateur astronomers. This software is tested to be sensitive in detecting meteors \cite{2012pimo.conf...44B}.

It is important to continuously monitor the activities of the meteoroid streams at various locations around the globe \cite{1997A&A...317..953J}. Although Europe and America are covered by large networks e.g. GMN and FRIPON, there is no extensive network in China. Enthusiasts in China have been operating meteor cameras for years, mostly under the name of China Meteor Monitoring Orgnization (CMMO, \url{http://www.qd-sky.cn/custom2561102.html)}. However, orbits of the meteoroid are rarely obtained due to a lack of homegrown tool-sets.

Commercial off-the-shelf CMOS cameras are crucial for the success of those meteor networks due to their low cost, wide availability, and high sensitivity\cite{2018pimo.conf...71S}. The CMOS cameras usually provide encoded video streams mostly in H.264 encoding \cite{1218189}, e.g. GMN \cite{2021MNRAS.506.5046V}. However, such encoding is a detrimental compression, which has negative impact on meteor detection and photometry, which will be demostrated in Section \ref{Evaluations}. The RMS software used by GMN uses Four-frame Temporal Pixel (FTP) format \cite{2011Icar..216...40J} specifically designed for storing image data of meteors. This format is heavily compressed when 4 frames represent a 256-frame-block, and only small rectangles containing the meteor is saved. A secondary detection procedure is used aiming to overcome the limits of the FTP compression. This limits the implementation of such pipeline, since other objects of interest (satellites, aircrafts, sprites etc.)  may be poorly preserved by the compression algorithm.

Furthermore, the timing precision in these cameras is not high enough to accurately determine the orbits of meteoroids \cite{2020MNRAS.491.2688V}. Most CMOS sensors used by meteor cameras work in rolling shutter mode, which means that the rows of the chip array are not exposed with the same time. This feature can cause ``jello effect" for fast-moving objects, which affects the determination of the meteor trajectory\cite{2018JIMO...46..154K}. It is fortunate that sensors with global shutters will ensure synchronous exposures across all pixels.

Nowadays, some CMOS cameras have built-in GPS modules, which provide hardware timestamps for video frames, e.g. QHY174 (\url{https://www.qhyccd.com/qhy174gps-imx174-scientific-cooled-camera/}). QHY174 is assembled with the Sony IMX174 sensor and works in global shutter mode. Although such kind of cameras are more expensive, it can achieve high-precision determination of meteoroid trajectories.

In order to detect faint meteors and obtain trajectories of corresponding meteoroids precisely, we designed a new meteor monitoring system with both improved spatial and temporal precision compared with other existing video-based meteor detection systems. A partner software is also developed.  Three motivations are driving the development of a new software:

\begin{enumerate}

\item[1.] Uncompressed data. We use USB CMOS cameras to capture the uncompressed video data and software to save FITS files containing uncompressed and uncropped images which enables detailed and targeted analysis for meteors and other types of objects.

\item[2.] Very short response time. The system is developed and optimized for reducing response time, whose ultimate goal is to calculate the landing locations of meteorites before landing.

\item[3.] Better dynamics. Aided by improved timing accuracy and positional accuracy, the dynamics of the meteoroid can be better observed. By accumulating the data in continuous operations, this system can test finer dynamics models of the meteoroid as they interacts with the atmosphere.

\end{enumerate}

The system can be easily extended to construct a large network covering China and surrounding areas, which will contribute to the continuous monitoring of meteoroid activities. By accumulating a large number of meteor events, this system can also help understand the properties of established meteoroid streams, discover new meteoroid streams, and perform the statistical study of the properties and origins of interplanetary dust particles.

This paper is organized as follows. Section \ref{sec:system} introduces the detailed design of our M$^3$ system. Section \ref{sec:software} describes the software design and details data processing algorithms. Section \ref{sec:test} shows some on-site testings and preliminary results. The summary and some further considerations are given in Section \ref{sec:summary}.

\section{Design and Components of the $\rm{M^3}$ system} \label{sec:system}
\begin{figure*}[htb]
    \centering
    \includegraphics[width=0.9\textwidth]{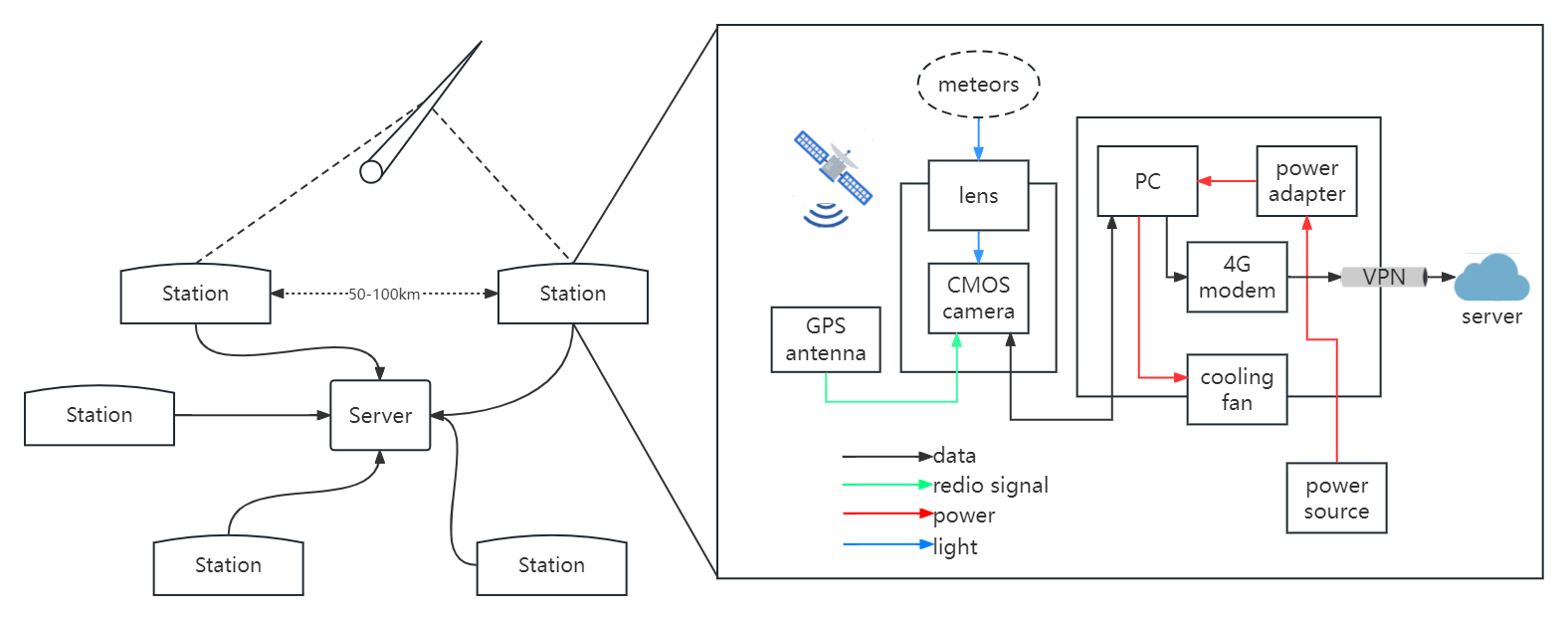}
    \caption{Framework of a meteor monitoring network (left) and the design of the M$^3$ system (right). The network has a central server and several stations, which are 50-100 km apart. Each station contains both camera and one instrument compartment. The camera compartment houses a combination of a camera and lens , which is connected to a computer in the instrument compartment. These devices are assembled with waterproof casing, which also has internet communication, cooling and power supplementation capabilities.}\label{hardware}
\end{figure*}

Figure \ref{hardware} shows the framework of a meteor monitoring network and our design of the M$^3$ system. The M$^3$ systems can be deployed at different sites, which are called ``stations". Ideally, these stations are apart from each other at a distance of 50-100 km, ensuring good trigonometric parallax measurements of detected meteors. Optimally, the network will be comprised of multiple stations uniformly distributed over a large area. These stations should be distributed as widely as possible to reduce the total cost and meanwhile be close enough to keep favourable trigonometric measurements. Each station is designed to run independently with local power supplies and computing capabilities. All the stations will be connected to a central server for gathering all the observed data.

The right panel of Figure \ref{hardware} presents the overall design of the M$^3$ system. It is designed to capture videos using wide field-of-view cameras and process the data in real time, and communicate with the central server for further processing. A local computer is required to finish such tasks. Furthermore, the system should be waterproof so that it can be installed outdoors. The design of the system includes two main parts: the camera and instrument compartments. The camera is equipped with a lens and a monochrome CMOS detector. It is embedded with a GPS module to provide timing signals. The instrument compartment mainly contains a computer and other electronics. Cooling fans are installed here for heat dissipation in the summer. For sites without internet connection, a 4G modem and VPN software can be installed for network communications. 

\subsection{Camera and Lens} 
Our meteor monitoring system has one or multiple ``optics+camera" unit, which is the combination of a wide field-of-view (FoV) lens and a high-performance CMOS camera. The camera module used in our system is QHY174GPS (\url{https://www.qhyccd.com/qhy174gps-imx174-scientific-cooled-camera/}) . It is a monochrome CMOS camera equipped with a SONY IMX174 sensor. The CMOS sensor has a size of 1,920$\times$1,200 pixels with a pixel size of 5.86 {\textmu}m, providing an effective image area of 11.25 mm $\times$ 7.03 mm. It has a QE of 78\% and low readout noise of 3-5 e$^{-}$, which is suitable for detecting faint meteors. The camera is connected to a computer using a USB3.0 connection, supporting continuous operation at 30 frames per second (FPS).

An important reason of choosing QHY174GPS is that it has a built-in GPS module that can be in sync with the atomic clock signals received from GPS satellites. It can record the start and end of the exposure time with $\sim$1\textmu s precision \cite{2023PASP..135b5001K}, which is beneficial for determining the meteor orbits and predicting the point of fall accurately with observations from different sites. The camera has a global shutter to ensure the same exposure time across the entire sensor. This feature is beneficial to detecting high-speed moving meteors. 

Limited by optical designs, fish-eye lenses usually do not have big apertures and hence prevent us from detecting faint meteors. The size of the sensor IMX174 is larger than typical meteor cameras used by other meteor networks (e.g. IMX291 by GMN), so lenses with a longer focal length can be mounted to achieve fainter limiting magnitudes with similar FoVs. In addition, longer focal length also increases astrometric accuracy of the position of the meteors.

However, the focal length of the lens cannot increase indefinitely, as the streak of a fireball can exceeds the FoV of the cameras, and images will be too saturated to do astrometric measurements. To balance the requirement for bright and faint meteors, lens from medium angle to wide angle are optimal for this system. Hence, we use AZURE-0614MLM (\url{https://www.azurephotonics.com/productinfo/249417.html}) from Azure Photonics. It has a small focal ratio of F1.4 and a focal length of 6.5 mm. When mounted with QHY174GPS, it provides a FoV of $88\times 58$ deg$^{2}$ and generates a stellar limiting magnitude of about 4.8 mag at the frame rate of 30 FPS.

\subsection{Computer}

The computer is used for controlling the camera to acquire the video data automatically, processing the video stream to identify atmospheric entry events in real time, calculating the coordinates, and managing the data and transferring them to the central server. The main considerations are as follows: 
\begin{enumerate}
\item[1.] To run our detection software (see Section \ref{sec:software}), at least four threads per camera are needed. Considering the workload from other tasks (e.g., displaying and uploading), we require a single-camera station to use a 4-core CPU, and a two-camera station to use a 8-core CPU.

\item[2.] The RAM usage depends on the resolution of the camera. About 100-200 frames are temporally loaded in RAM for meteor detection. For image size of the QHY174GPS, about 1 GB RAM is needed for each camera, and additional 2 GB for each station is required for the buffer storage when processing meteor data and writing them to hard disks. 

\item[3.] Each station is equipped with SSD hard drives to facilitate data storage. The fast read and write speeds ensure the system can collect and process the data in time for real-time observations. During operation, the raw and processed data will be automatically compressed and uploaded to the central server. Once uploaded, the local files can be deleted so it does not clog the computers' storage. However, for stations that lack proper networking, a larger hard drive will need to be installed. The size of one frame of the camera is 2.3 MB, and assuming 50 frame per meteor and 20 meteors per night, it is estimated that each camera will generate about 50 GB of data in one month.

\item[4.] The VPN software installed on the computer is used to transfer data and receive commands via Ethernet or 4G communication depending on the location of the station. It enables remote controlling through SSH or remote desktop and troubleshooting especially during the initial phases after installation.
\end{enumerate}

Our chosen computer is equipped with an Intel i5-8250 CPU, 8 GB RAM, and a 120 GB SSD disk. It is powered by an external power adapter providing 19V DC current. The computer case is compact enough to fit the waterproof casing as described in Section \ref{sec:casing}.

\subsection{Waterproof Casing} \label{sec:casing}

Because the connection distance between the camera and computer is limited by the maximum length of USB3.0 cables (3-5 m), these devices are protected by a water-proofing casing that will be placed outdoors. We specially design a waterproof casing. It consists of two main compartments: the camera and instrument compartments. The camera compartment houses two mounting points that can be tilted vertically to point the cameras to a specific sky region. Optical windows made of quartz protect the cameras and lens. A GPS antenna for the QHY174GPS camera is attached on the top side of the casing by magnets. The instrument compartment houses the computer, cooling fans, power supply and a 4G network card, which are needed if the observation site has no internet connection. Two plastic widows are specially opened on each side of the casing for 4G signals to pass through. The cooling fans and corresponding air vents are arranged for cooling all devices in the instrument compartment in the summer. Furthermore, we also reserve space for batteries if solar power is used.

\begin{figure*}[htb!]
    \centering
    \includegraphics[width=0.8\textwidth]{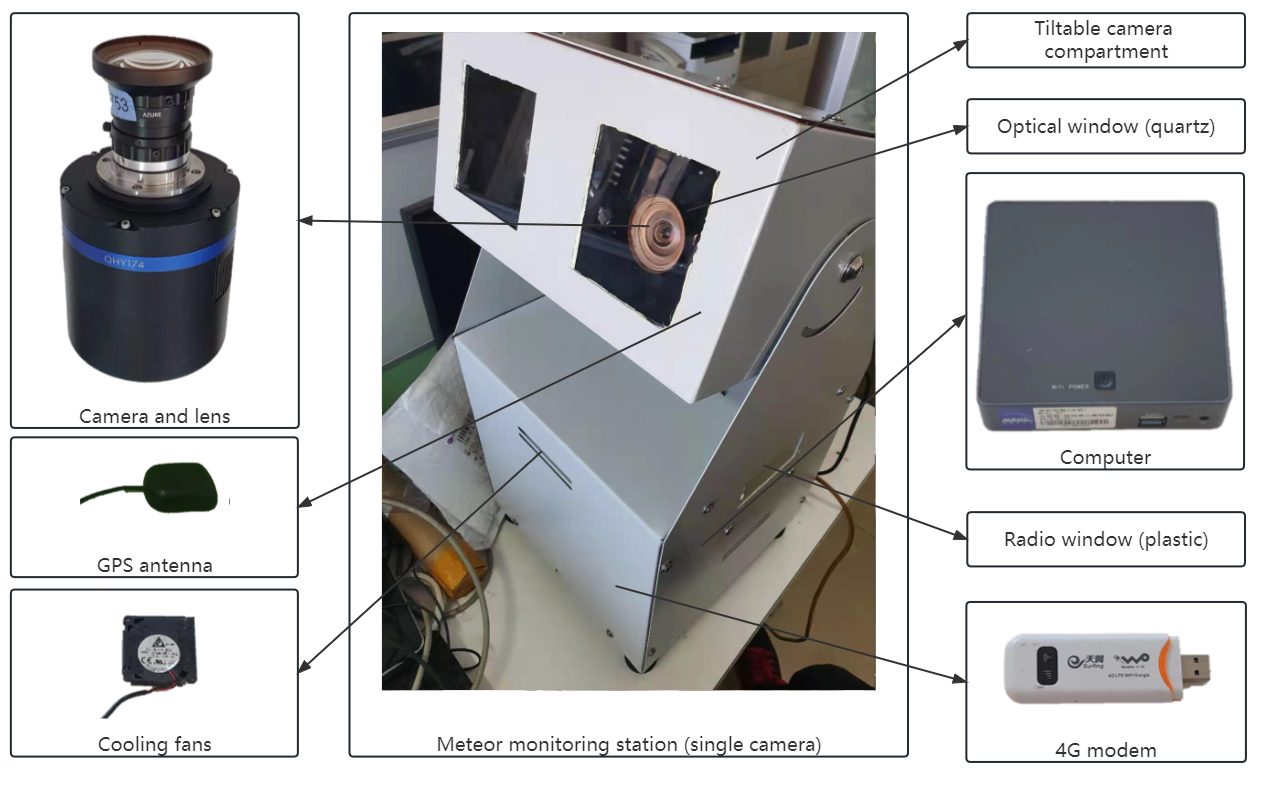}
    \caption{Components of a meteor monitoring system.}\label{hardfull}
\end{figure*}

The whole station weighs about 10 kg and consumes about 30 W power when the meteor detection software is running. The detailed parameters of each component are listed in Table \ref{spec_table}. The main cost of each station comes from the camera, lens and computer. The total hardware cost is about \$2,000.  Figure \ref{hardfull} presents the photos of each device and the final assembly.

\begin{table*}[htb!]
\centering
\caption{Parameters of the M$^3$ system.}
\label{spec_table}
\begin{tabular}{@{}lll}
\hline
\hline
Category & Item & Value\\
\hline
QHY174GPS camera&CMOS chip&SONY IMX174 CMOS\\
&resolution&1,920$\times$1,200\\
&pixel size&5.84 \textmu m\\
&quantum efficiency&78\%\\
&shutter&global\\
&bit depth&8/12 bit\\
&read out noise&3-5 e$^{-}$\\
AZURE-0614MLM lens&focal length&6.5 mm\\
&F No.&F1.4\\
&image circle&1 inch\\
&FoV&$88^\circ\times58^\circ$\\
&limiting magnitude&4.25 m\\
Camera control&expo. time&33 ms\\
&frame rate&30 FPS\\
&gain setting&48\\
&offset&500\\
computer hardware&CPU&intel i5-8250U\\
&RAM&8 GB\\
&storage&120 GB SSD\\
&power consumption&peak 30 W\\
\hline
\end{tabular}
\end{table*}

\section{Meteor detection and measurement software} \label{sec:software}
\subsection{Overview of the Software}
\label{sub-soft}
\begin{figure*}[!htb]
    \centering
    \includegraphics[width=\textwidth]{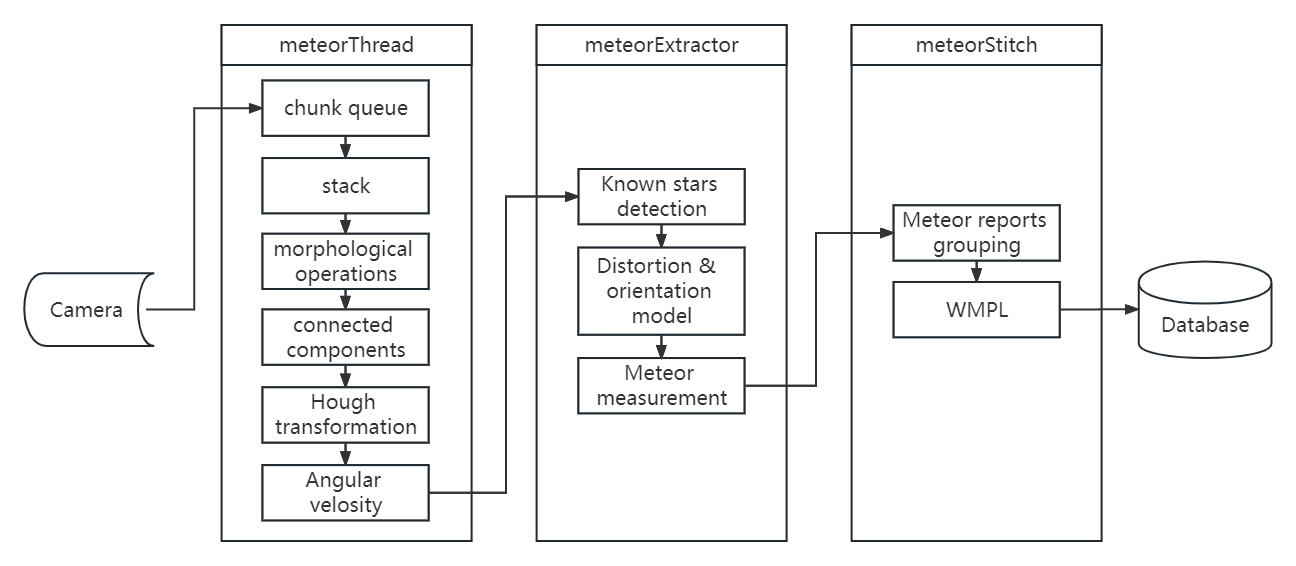}
    \caption{General flowchart of our meteor monitoring software.}
    \label{softwares}
\end{figure*}

From a single station, we can only obtain the two-dimensional coordinates of meteors. To determine the 3-D trajectory of a meteor through triangulation, at least two stations are required. Our meteor monitoring software includes following main processes: (1) identifying meteors from the CMOS video stream (\textit{meteorThread}); (2) measuring the coordinates of detected meteors (\textit{meteorExtract}); (3)  determining the orbits using coordinates from multiple stations (\textit{meteorStitch}). Figure \ref{softwares} presents the general flowchart of the software.

\begin{itemize}
\item[1.] \textit{meteorThread} is responsible for capturing meteors from the camera video. Generally, a few tens of meteors could be detected through each clear night. It is not necessary to store the full raw video data. A meteor detecting algorithm as shown in Section \ref{sec:identify} is developed to recognize the meteors from the video. \textit{meteorThread} controls the cameras and processes the video data in real-time. It only saves the video clips with meteors and discards the data with no moving objects. Sometimes, airplanes or satellites can be detected, which are distinguished from the meteors according to their moving velocity, as meteors move much faster. The difference of the angular velocities of meteors and other objects are analyzed in Section \ref{sec:identify}.

\item[2.] \textit{meteorExtract} is to position the meteors in the celestial coordinate system. It measures the pixel coordinates of both meteors and stars in each video frame. A coordinate transformation is determined by using the stars with known celestial coordinates as the reference, which will be described in Section \ref{sec:position}. The transformation is used to convert the pixel coordinates to the equatorial coordinate system for the objects detected in each frame. At the same time, the precise timestamp of the meteor is derived from the records of the GPS module. All of this information will be compiled into a report file. The video clips together with the report files will be synchronized to the central server.

\item[3.] \textit{meteorStitch} is to calculate the spatial trajectory of each meteor by combining at least two records of the same meteor from different stations. It monitors the catalogs in the central server and identifies data belonging to the same meteor events according to the coordinates and timestamps. A open-source Python package of \textit{WesternMeteorPyLib }(\textit{WMPL}) \cite{2020MNRAS.491.2688V} is adopted to calculate the meteor trajectories and determine the orbits of the meteoroids in the solar system. 

\end{itemize}

\subsection{Meteor detecting algorithm}
\label{sec:identify}
\begin{figure*}
    \centering
    \includegraphics[scale=0.45]{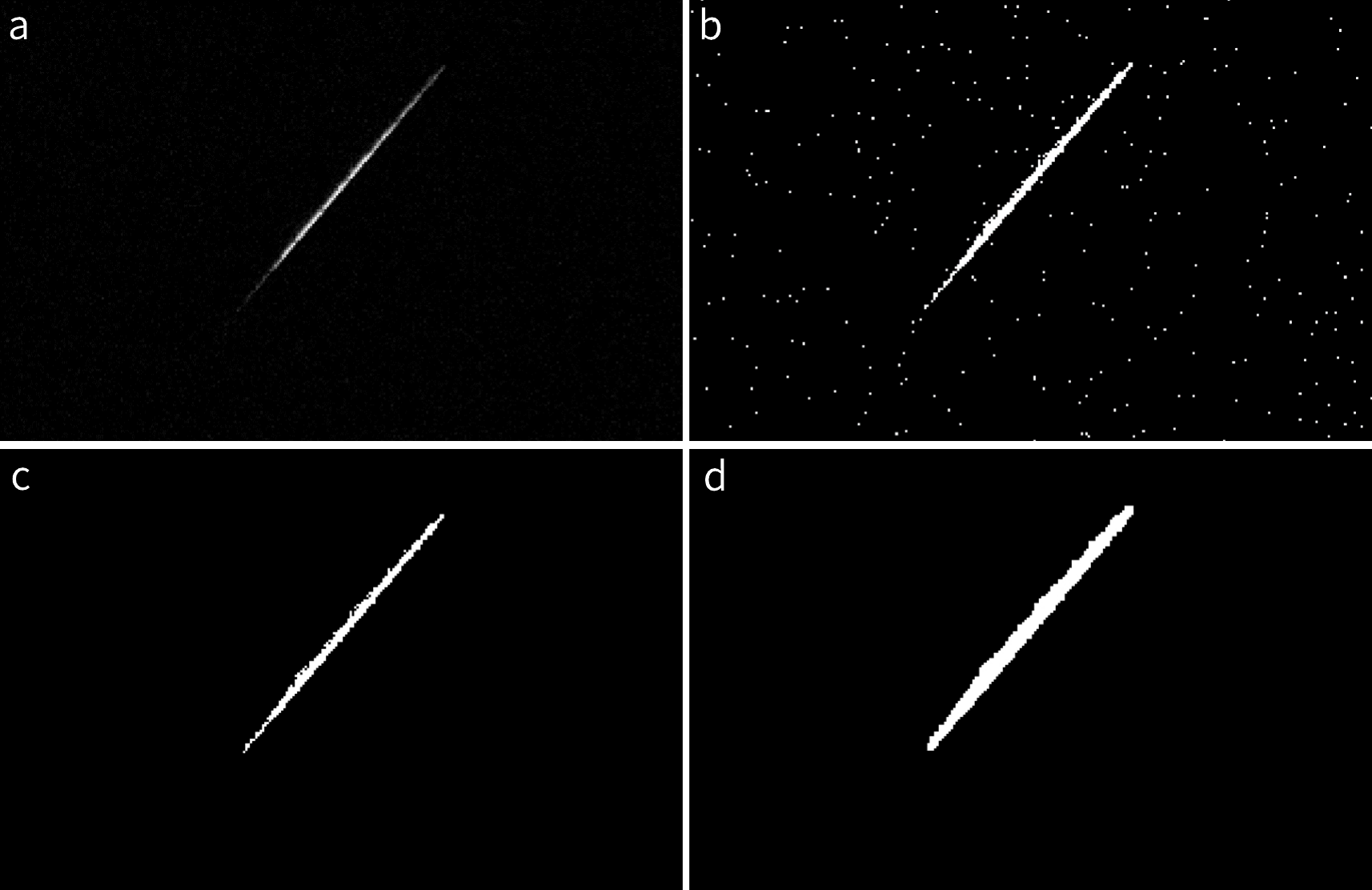}
    \caption{(a) Stacked image of 50 consecutive frames to identify meteors. (b) Binarized image of the stacked image. (c) The binary image with noise pixels removed. (d) Dilated binary image.}
    \label{fig:detect}
\end{figure*}

Because of the requirement of the accuracy and efficiency of detecting meteors from real-time video data, it can be quite a challenge to design the detecting algorithm. Generally, the transient objects can be easily identified in consecutive frames, because their brightness or positions shift. Meteors are moving objects which present linear trajectories in the video or emergent line morphologies in the stacked image created by consecutive frames, so they can be identified as straight lines \cite{2016pimo.conf..307V,suipian2009}. We adopt a similar method of detecting meteors in this paper, which is described as below:

\begin{enumerate}
    \item [1.] The video is divided into small segments called chunks, each with an duration of around 2 seconds ($\sim$ 50 frames). Each video chunk is stacked into a maxima image by taking the maximum of each pixel in all 50 frames.
    \item [2.] The maxima image is subtracted by the average of two neighbour stacked maxima images. Any moving objects will appear as line trails in the residual image as shown in Figure \ref{fig:detect}a. The residual image is binarized (see Figure \ref{fig:detect}b), where pixels belonging to celestial objects present 1 and the background pixels present 0. The threshold for discriminating the background from object signals is set to be 2.5 times the standard deviation above the average background. 
    \item [3.] Noise points are removed as shown in Figure \ref{fig:detect}c if the total number of pixels (i.e. area) of an object are smaller than 5. The threshold is specified by the configuration, which is empirically determined by the FWHM of stars. The binary image of remaining objects are dilated as shown in Figure \ref{fig:detect}d, in case that some meteor trails are disconnected due to the brightness variation or frame loss of the CMOS camera. 
    \item [4.] The trails are identified using the Hough transformation. If moving objects are detected in two consecutive video chunks with similar Hough parameters, they are regarded as the same object. Very bright meteors which appear as elongated blobs in the images, can also have meaningful Hough transformation results which indicating their moving directions. Their position, moving direction and speed will be measured for further object classification. 
\end{enumerate}
The thresholds mentioned above are adjustable parameters and may be modified according to the local observing conditions. 

Not all moving objects are meteors. Some of them may be airplanes and artificial satellites. However, these man-made objects move much slower and typically travel across the whole night sky. If the moving object appears in only one video chunk, it is always considered as a meteor, since satellites and airplanes stay visible during their traveling across the FoV. For those objects appear in several chunks, their angular velocities are roughly calculated according to the GPS time stamps of the chunks. We mainly use angular velocities to rule out satellites and airplanes, the threshold of which is set to be the highest possible angular velocities. The most extreme condition for satellites is a highly elliptical orbit with a low perigee at the zenith of the station. Such satellites can move as fast as 11.2 km/s, assuming that the perigee is at 200 km. This gives the maximum angular velocity of $3.2 ^{\circ}$/s. Commercial airliners at $\sim$10 km cruising altitude fly at $\sim$1000 km/h, and it gives $1.6 ^{\circ}$/s. Airliners during take-off and landing are much closer to the ground and may have angular velocities up to $7 ^{\circ}$/s, but usually we try to avoid installing the meteor monitoring devices near busy airports.

The above meteor detecting process is implemented with a multi-thread program. One thread is used to control the camera for gathering the video data and generating video chunks. The chunks are organized in a queue and fed to another thread, which is designed to identify the moving objects and to distinguish the meteor from man-made objects. A third thread is to display the real time frames on the screen. The forth thread is to save the data of detected objects into a multi-extension FITS file, which includes the raw video frames, the related GPS timestamps and the detection information for each meteor. The video data for non-meteors may be deleted in order to save the storage space.  

\subsection{Meteor positioning algorithm}
\label{sec:position}
\begin{figure*}
    \centering
    \includegraphics[width=0.8\textwidth]{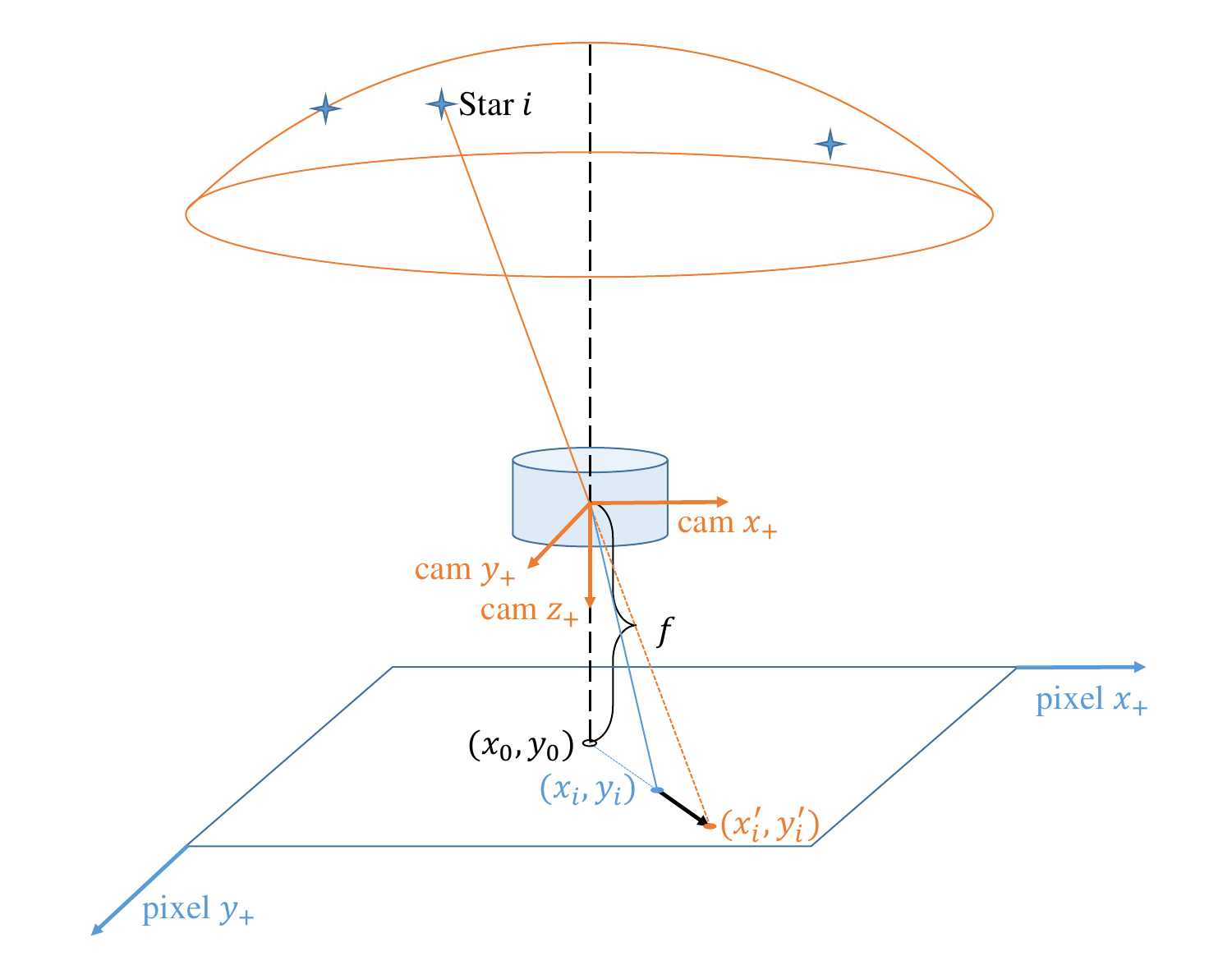}
    \caption{Conversion from pixel coordinates and camera coordinates. A perfect lens without optical distortion project a star from the celestial sphere onto the focal plane at ($x^\prime_{i}, y^\prime_{i}$). However, the distance between the real pixel coordinate and optical center has been distorted by the camera (black arrow). The actual coordinate is at ($x_i$, $y_i$). The distortion can be described with a even-order polynomial.}
    \label{distor}
\end{figure*}

To obtain an astrometric solution for converting pixel coordinates on the CMOS chip into celestial coordinates, we measure the positions of stars in individual frame. The astrometric solution includes both a linear transformation and a distortion. We adopt a mathematical derivative of the Brown-Conrady distortion model \cite{1971Close} to describe the non-linear change of the pixel coordinates in relation to angular distances between two objects, which is illustrated in Figure \ref{distor}. It is modelled as a rotational symmetric polynomial for our camera lenses:

\begin{equation}
\begin{split}
\left(
\begin{array}{l}
x^\prime_{i} \\
y^\prime_{i}
\end{array}
\right)
=
\left(
\begin{array}{l}
x_i-x_0 \\
y_i-y_0
\end{array}
\right)
(1+k_1r^2+k_2r^4+\cdots), \label{equ1}
\end{split}
\end{equation}
where ($x_i$, $y_i$) is the pixel coordinate of a specified star, ($x_0$, $y_0$) is the pixel coordinate of the camera optical center, ($x^\prime_{i}$, $y^\prime_{i}$) is the virtual undistorted coordinate, and $r=\sqrt{(x_i-x_0)^2+(y_i-y_0)^2}$ is the radial distance. 

The camera coordinate ($x^\prime_{i}$, $y^\prime_{i}$, $f$) can then be converted to the celestial coordinate ($x_{ci}$, $y_{ci}, z_{ci}$) using a rotational transformation matrix:

\begin{equation}
\left(
\begin{array}{l}
x_{ci} \\
y_{ci} \\
z_{ci}
\end{array}
\right)
=\mathbf{R}
\left(
\begin{array}{l}
x_i^{\prime} \\
y_i^{\prime} \\
f
\end{array}
\right), \label{equ2}
\end{equation}
where $\mathbf{R}$ is the rotational matrix, $f$ is the focal length of the camera, and ($x_{ci}$, $y_{ci}$, $z_{ci}$) is three-dimensional spherical coordinate, equivalent to the equatorial coordinate ($\alpha$, $\delta$). 

Generally a even-order polynomial is enough to describe the distortion of our lenses. In this case, there are 8 parameters to be determined in both Equations (\ref{equ1}) and (\ref{equ2}), including $x_0$, $y_0$, $f$, $k_1$ and $k_2$, and 3 degrees of freedom in $\mathbf{R}$. The 8 parameters are determined using the Nelder–Mead method \cite{1964A}, implemented by \textit{SciPy} \cite{2020SciPy-NMeth}. This algorithm searches the optimized parameter to minimize residuals from all the inputs, which are the pixel and celestial coordinates of known stars.

Considering external factors such as strong wind, ambient temperature variation, and artificial interference that affect the cameras' status, the 8 parameters are determined in real time using stars for each detection, providing a ``on-line" calibration. This calibration method can be fully automated  by \textit{meteorExtract}, once initial parameters are provided. The pixel coordinates of stars are measured from the average of the video clip consisting of $\sim$50 frames. During this period, the stars are considered stationary. 

At the setup stage, we first capture an image of the sky, and then manually select each visible star in the image. The software will perform an preliminary calibration using the measured pixel coordinates and celestial coordinates from the HIPPARCOS Catalogue \cite{1997A&A...323L..49P}. The calculation of the coordinates at J2000 epoch is facilitated by \textit{Astropy} \cite{2022ApJ...935..167A} and \textit{Skyfield}  \cite{2019ascl.soft07024R}. At the stage of regular operations, the software can automatically match the stars in images with those in the HIPPARCOS Catalogue. Bright stars are used, and the matching radius is set to 3 degrees. An upper limit of the number of stars (15 in our practice) is set to only include the bright stars that are far away from each other and not prone to be miss-matched. In most cases, the stars brighter than 2.5 mag are used which can result in astrometric solutions with enough accuracy, as shown in Section \ref{Astrometric accuracy}. The limited number of stars also ensures fast convergence of the solutions, as we aim to limit the time cost for astrometry to less than 5 seconds.

Although meteors appear as long linear lines in stacked images, they are much shorter ($\sim$10 pixels) in individual frames, so we determine their positions by calculating the centroids. The background image is calculated by averaging the pixels without the meteor. Then each frame is subtracted by the background to get the net fluxes of the meteor pixels. The positions of the meteors are determined by the fluxes of the pixels within the detected connecting component. The celestial coordinates of meteors are calculated by using the above astrometric solution.  

\subsection{Orbit determination from multi-station observations}
The central server collects the coordinates reported by all M$^3$ systems deployed at different stations and calculates the 3-D trajectories of meteoroids using the multi-station data. Firstly, we need to group the data from different stations that belong to the same meteoroid. The camera has a built-in GPS that can provide accurate entry time for each meteor. Usually, it is enough to group the meteors according to their entry times. However, there are some situations when multiple meteors could emerge at the same time (e.g. a meteor shower). In this case, we check whether two lines of sight of the meteors observed at the two stations intersect or not. If these two lines intersect, we consider that the meteors detected at two stations are from the same meteoroid.

Secondly, we utilize the \textit{WMPL} to determine the orbit of each meteor using the grouped data. The \textit{WMPL} takes the equatorial coordinates of a meteor observed from multiple stations and locations of these stations and then calculates the 3D trajectory by a novel method of trajectory estimation \cite{2020MNRAS.491.2688V}. In our software, the equatorial coordinates are converted to the current epoch, and the height of the stations are converted from WGS84 to mean sea level height as required by the \textit{WMPL}. The information is provided to the \textit{WMPL} for determining the orbit of the meteors.

The processes of data grouping and orbit determination are wrapped into a program \textit{meteorStitch}. This program collects images and coordinates sent by all M$^3$ systems. When new reports from a single station are received, it checks with existing reports in the database for grouping. If a new report does not belong to any existing groups, it will be assigned to a new group. Otherwise, it will be merged into an existing group and the \textit{WMPL} will be triggered to update the orbit of the group. The orbit parameters along with the detection information will be stored in the database for future statistical investigations. By using the \textit{WMPL} Monte-Carlo solver, we can obtain the 3-D trajectory, velocity, radiant, and their uncertainties. The solver can also provide heliocentric orbit elements based on the velocity estimate. 

\section{On-site testing and preliminary results} \label{sec:test}
\subsection{Monitoring the Geminid meteor shower}
We select major meteor showers to test the performance of our system. The Geminid meteor shower is one of the most high-profile meteor showers occurring annually. It has been reliably predicted to put out a large number of meteors within several nights. According to the IAU meteor shower catalog (\url{https://www.ta3.sk/IAUC22DB/MDC2022/Roje/pojedynczy_obiekt.php?lporz=00028&kodstrumienia=00004&colecimy=0&kodmin=00001&kodmax=01222&lpmin=00001&lpmax=01711&sortowanie=0}) and the 2021 Meteor Shower Calendar published by the International Meteor Organization (\url{https://www.imo.net/files/meteor-shower/cal2021.pdf}), the Geminid meteor shower reached its maximum at December 14th 2021, 07$^{\rm h}$ UT ($\lambda_\odot=262.2^\circ$). The radiant drifts from $\alpha = 108^{\circ}, \delta=+33^{\circ}$ to  $\alpha = 113^{\circ}$, $\delta=+33^{\circ}$ during December 10-15. The velocities at infinity ($V_{\infty}$) of the meteoroids are $\sim$35 ${\rm kms^{-1}}$.

During the Geminid meteor shower in 2021, we conducted a test observation with our meteor monitoring systems. Two identical sets of M$^3$ stations were deployed at two locations (denoted as ``A" and ``B") in the north of Beijing, which are 55 km apart. The observations started on December 11 and ended on December 14. During the 4-night observations, we detected 818 meteors at Station A and 723 ones at Station B.

\subsection{Astrometric accuracy}\label{Astrometric accuracy}

\begin{figure*}[tbh]
    \centering
    \includegraphics[width=\linewidth]{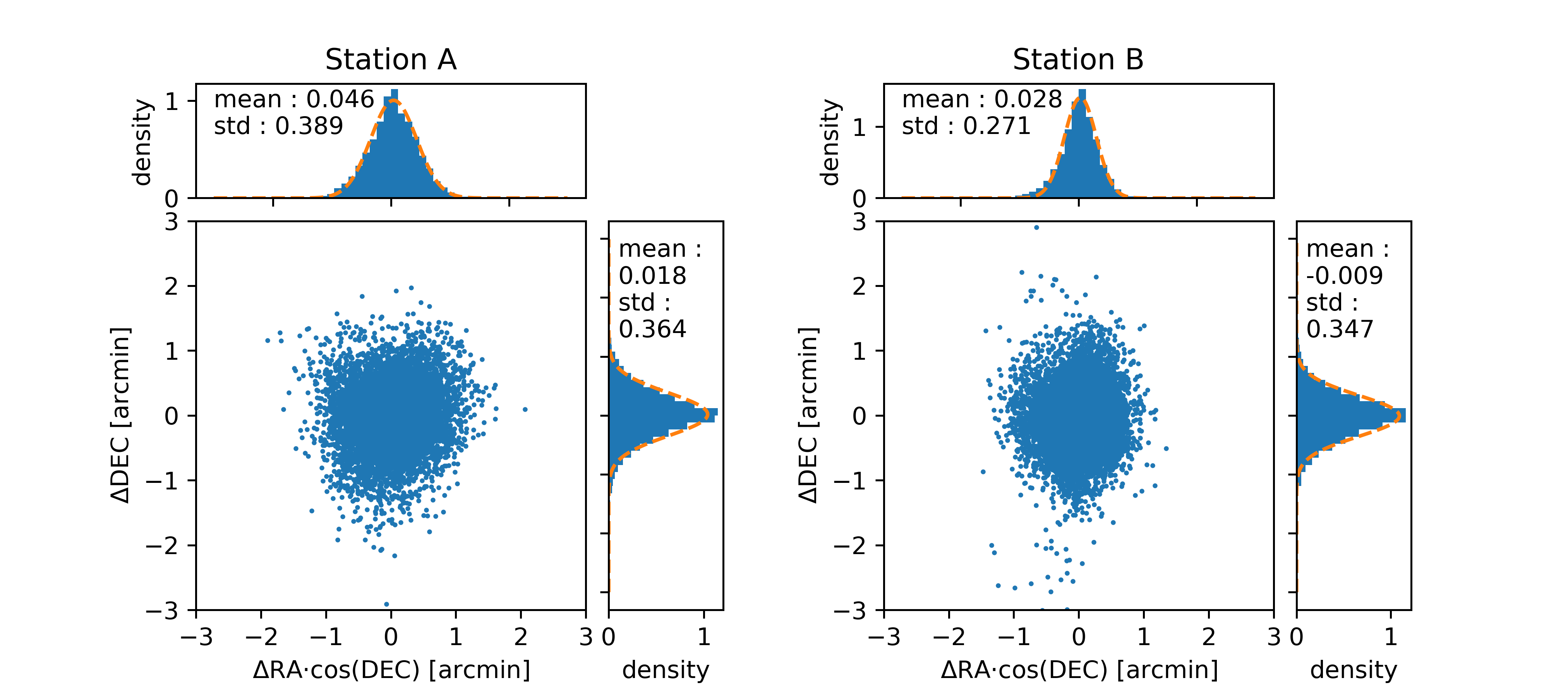}
    \caption{The distributions of the astrometric residuals of the position of the stars for two M$^3$ systems deployed at two stations (left for Station A and right for Station B). The top and right panels in each plot are the histograms of the astrometric residuals in R.A. and decl., respectively. The orange dashed curve in each panel is the best fitted Gaussian profile, whose mean and standard deviation are also marked.}
    \label{fig:resi}
\end{figure*}

To obtain the equatorial coordinates for the detected meteors, up to 15 brightest stars in each video clip are used to perform the astrometric calibration. The residuals between the measured coordinates and the ones in the reference catalog of the HIPPARCOS Catalogue \cite{1997A&A...323L..49P} are used to estimate the positioning accuracy. More than 1,000 stars have been detected in successive video frames and about 90\% of them have astrometric residuals smaller than 1 arcmin. Figure \ref{fig:resi} shows the astrometric residuals for the M$^3$ systems at two different sites. The positioning accuracies of Station A in R.A. and decl. are about 0.36 and 0.39 arcmin, respectively, while those of Station B are about 0.27 and 0.35 arcmin, respectively. Station B has a noticeably smaller residual in the R.A. direction, possibly caused by the imperfect assembly of the lens and camera. As the validation, in each video we randomly choose a star and calculate its celestial coordinates using the astrometric solution obtained by the other stars. The results that the residues between the calculated coordinates and the coordinates in the star catalog are similar show that there is no obvious over-fitting.

\begin{figure*}[!htb]
    \centering
    \includegraphics[scale=0.2]{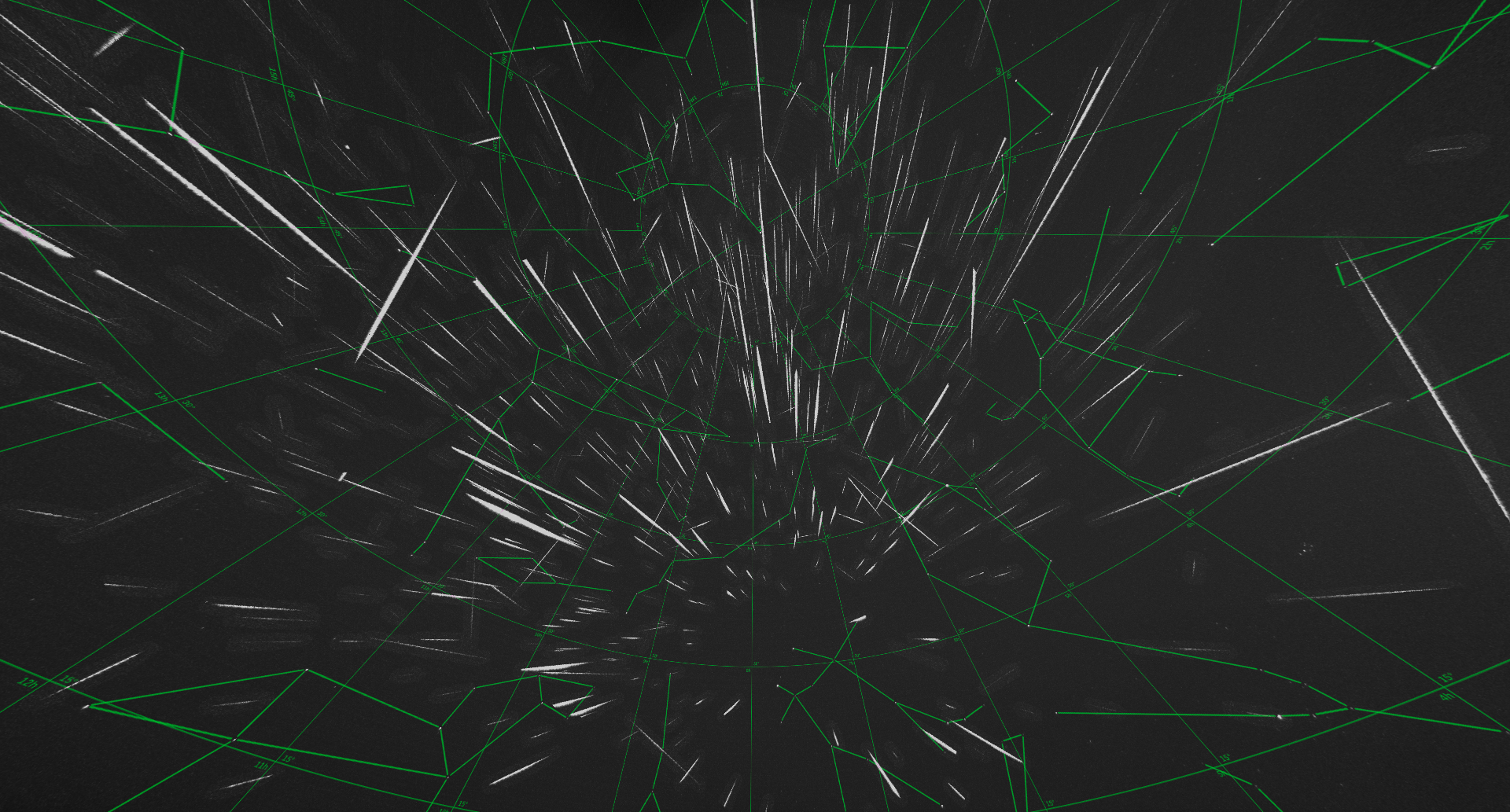}
    \caption{Panorama of the meteors detected at Station A. It includes all meteors detected in 4 nights. The bright stars (white spots) are connected by constellation lines in green.}
    \label{fig:pano}
\end{figure*}

Based on the astrometric solutions, we can project the images of the detected meteors into a common coordinate frame and stack the projected images into a panorama picture, which is shown in Figure \ref{fig:pano}. From this figure, we can clearly see the shower radiant. The bright stars connected by the constellation lines show no misalignment, indicating good astrometric accuracy.   

\subsection{Trajectories and orbits of the meteors}
\begin{figure}[htb!]
    \centering
    \includegraphics[scale=0.8]{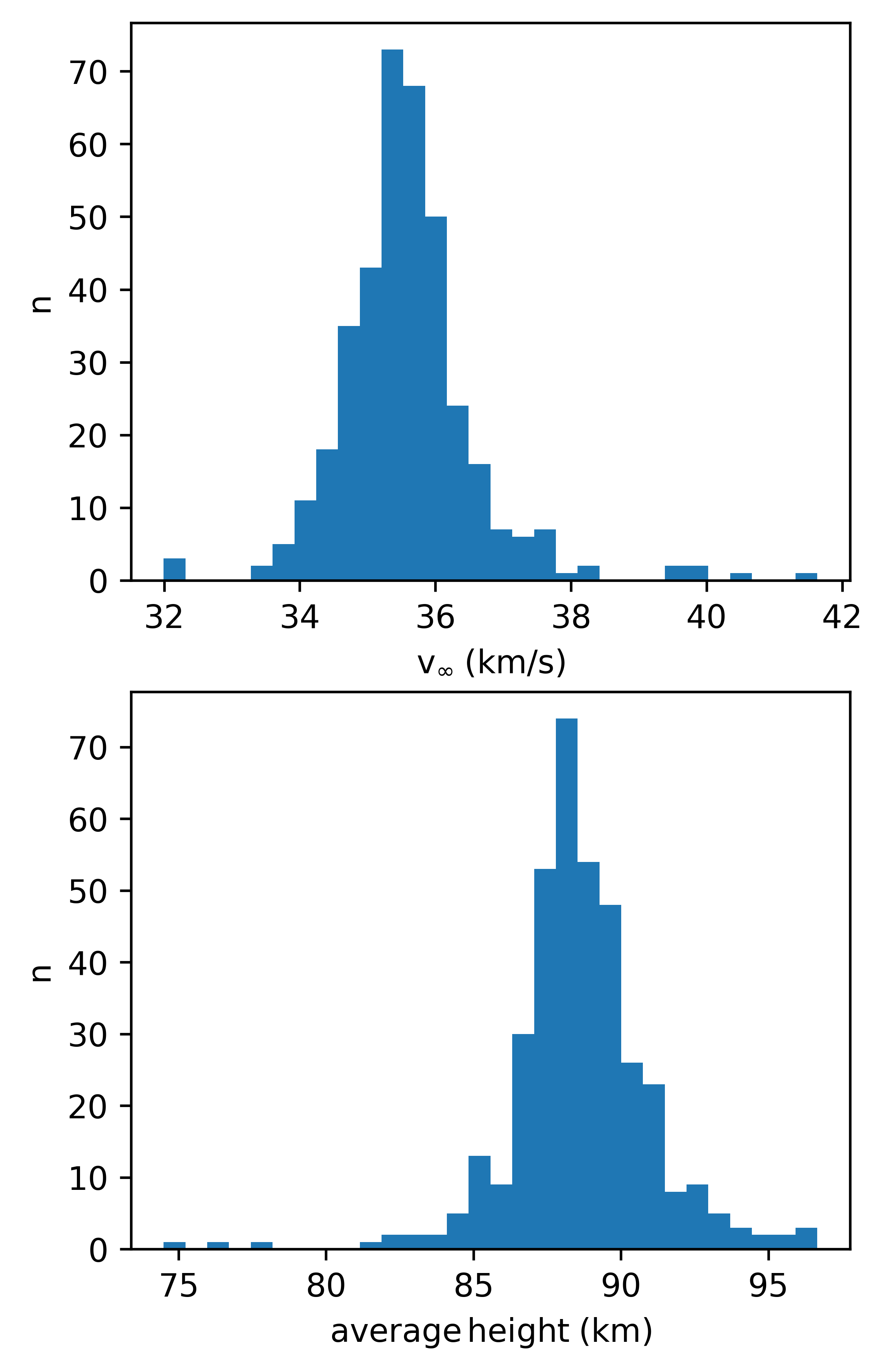}
    \caption{Distributions of altitudes (upper panel) and velocities of Geminid meteors (lower panel). }
    \label{speed}
\end{figure}

The spatial trajectories of the meteors detected simultaneously at two stations are determined by the \textit{WMPL}. A total of 473 meteor trajectories are solved successfully. Figure \ref{speed} shows the distributions of the average altitudes and velocities of these meteors. It is found that most meteors are located at 80-100 km in the upper atmosphere. The median altitude is 88.5 km. The velocity is about 33-38 ${\rm kms^{-1}}$ and the median velocity is 35.52 ${\rm kms^{-1}}$. The velocities and radiant of these meteors coincide with the IAU meteor shower catalog. In addition, it is notable to find that the time residuals of the triangulation fitting are mostly $\sim$1 $\mu$s, which corresponds to the advertised precision of the camera.

\begin{figure}[htb!]
    \centering
    \includegraphics[scale=0.15]{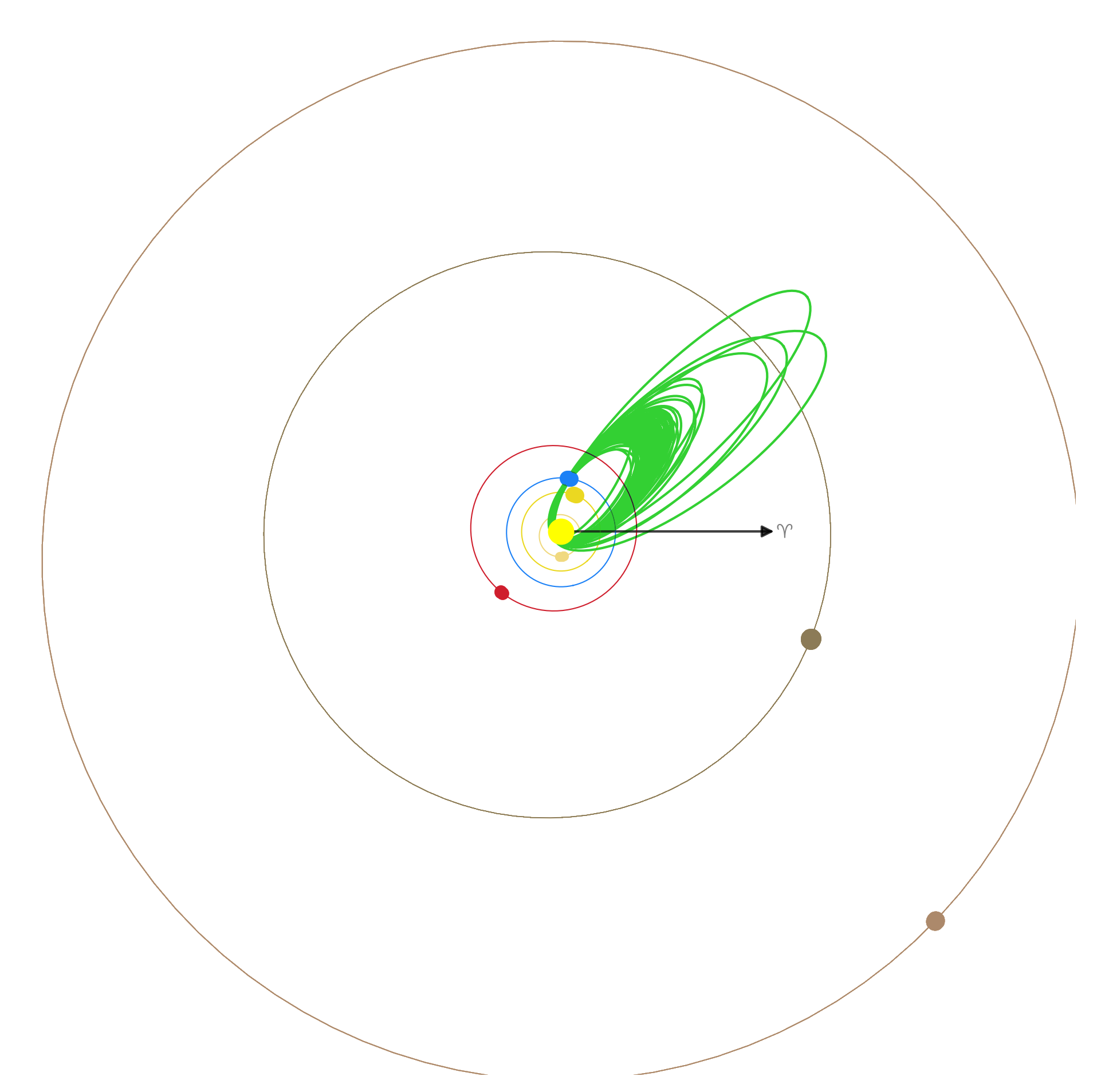}
    \caption{Heliocentric orbits of the observed Geminid meteoroids. }
    \label{heliocentric}
\end{figure}

Further looking at the radiant of each trajectory, 377 meteors are from the Geminid meteoroid stream. All the radiants are within 1 degree. With the estimate of the velocities at infinity, the \textit{WMPL} also calculates the orbital elements of the meteoroids before they enter the Earth's atmosphere. The orbits of these meteors in the solar system are presented in Figure \ref{heliocentric}.

The Geminid meteoroid stream has been well studied. It was reported that this stream is associated with asteroids (3200) Phaethon, 2005 UD, and 1999 YC \cite{2006mspc.book.....J, Jewitt_2006, Kasuga_2008}. The peculiar association with multiple asteroids indicates a violent catastrophic formation of the Geminid meteoroid stream, which is quite different from the cometary origin of other meteor streams. It is further supported by Ref.~\citenum{Cukier_2023} which used the data from the Parker Solar Probe and a dynamic model to show a catastrophic formation scenario of the Geminid meteoroid stream. We list the median orbital elements of our detected meteoroids in Table \ref{elements} and compare them with those of asteroid (3200) Phaethon. The orbit of the Geminid meteoroid stream is almost the same as the parent body of asteroid (3200) Phaethon, while the perihelion is slightly higher. Our observations thus coincide with the existing knowledge that the Genimid stream has an asteroid origin.

\begin{table*}
\centering
\caption{Orbital elements of our detected Geminid meteoroids}
\label{elements}
\begin{tabular}{@{}lll}
\hline
\hline
Orbital elements & Median of meteoroids & (3200) Phaethon\\
\hline
Eccentricity ($e$) & 0.887 & 0.88990\\
Semi-major axis ($a$) &1.30 AU & 1.27134 AU\\
Inclination ($i$)&23.1$^{\circ}$&22.2735$^{\circ}$\\
Longitude of the ascending node ($\Omega$) &262$^{\circ}$&265.172$^{\circ}$\\
Argument of perihelion ($\omega$) &324$^{\circ}$&322.216$^{\circ}$\\
Perihelion & 0.147 AU & 0.13998 AU\\
Aphelion & 2.45 AU & 2.4027 AU\\
\hline
\end{tabular}
\end{table*}

\subsection{Evaluations}\label{Evaluations}

The main purpose of video encoding is to reduce the size of video data while preserving as many details as possible that are important to human eyes. Thus, the details not perceivable to human eyes are removed during the compression, and the astronomical pipelines cannot get the correct results from the compressed data. Simulations were preformed to test the effectiveness of uncompressed data format. During each simulation, the detected image data are compressed to H.264 format with various bit-rate (8,000 kbps, 5,000 kbps and 2,000 kbps) to simulate the output of an IP camera, and the videos are uncompressed to replace the original data in the meteor data structures. The compressed data are then compared with the raw data in the perspective of photmetry and astrometry.

\subsubsection{Photometry}

\begin{figure*}[!htb]
    \centering
    \includegraphics[scale=0.9]{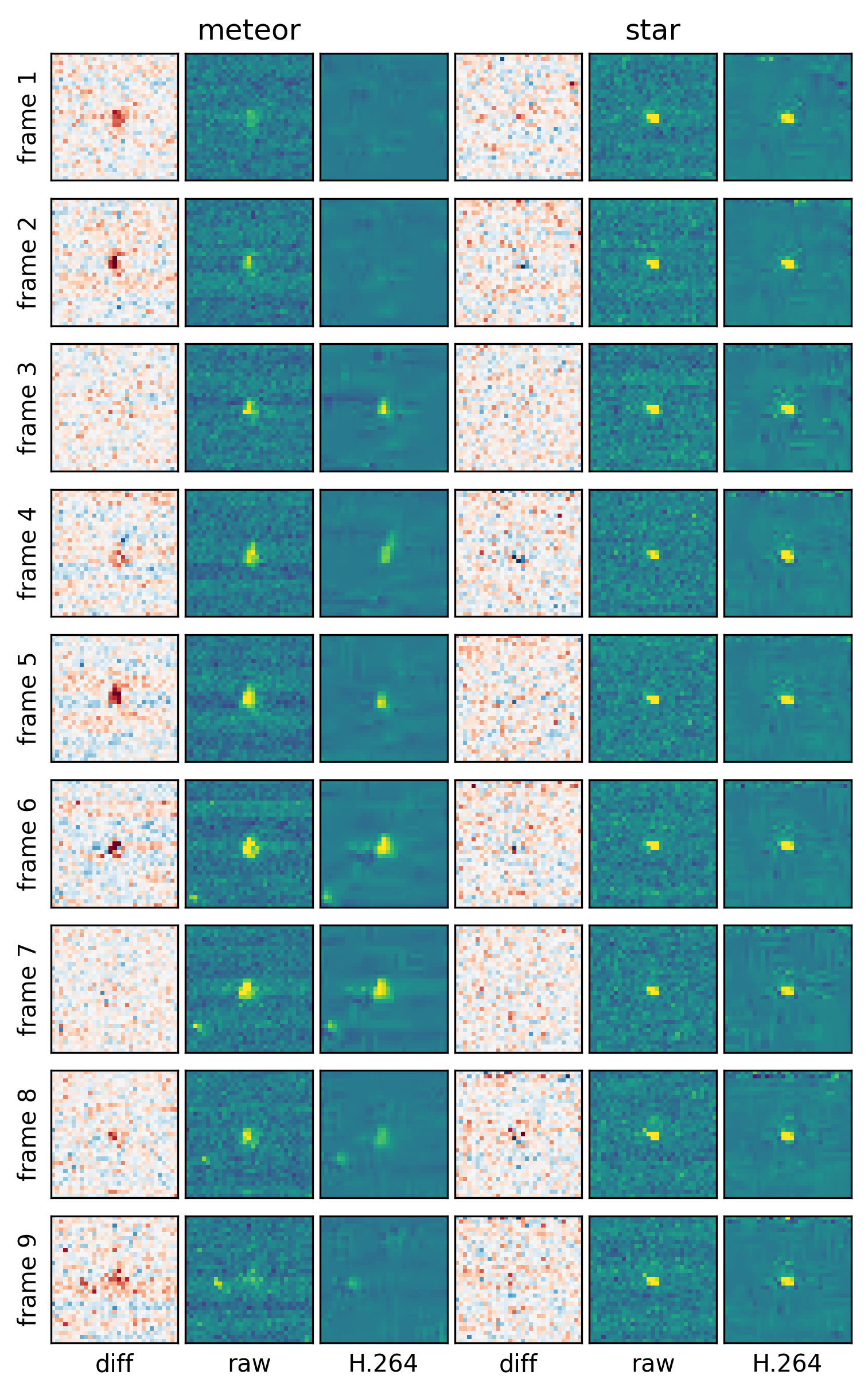}
    \caption{The comparison of the original frames and the compressed frames at 5,000 kbps bit-rate. The left half shows the image patches around the moving meteor and the right half shows the image patches of a star with similar brightness. Each row represents a frame from a continuous sequence where the patches of the meteor and the star are sampled. In the plots showing the differences, pixels with reduced values are colored red, and the ones with increased values are colored blue.}
    \label{fig:frame_compare}
\end{figure*}

Figure \ref{fig:frame_compare} shows the original frames and the compressed frames of a meteor and a star together with their differences. These frames are cropped to show the details around the moving meteor and the star. The image of the meteor and the star are affected differently by the compression. The difference of the star image remains mild and constant, while the difference of the meteor image varies significantly. In some of the frames (1, 2, 4, 5 and 6), the pixel values of the meteor images are underestimated, but frames 3 and 7 are affected much less severely. This behavior may be caused by the different frame types defined by the H.264 encoding \cite{1218189}.

Additionally, as shown in Figure \ref{fig:frame_compare}, much of the spatial and temporal noise of the sky background is removed during the compression. This may influence the background estimation and cause the overestimate of the SNR of the photometric results.

\begin{figure*}[!htb]
    \centering
    \includegraphics[scale=0.6]{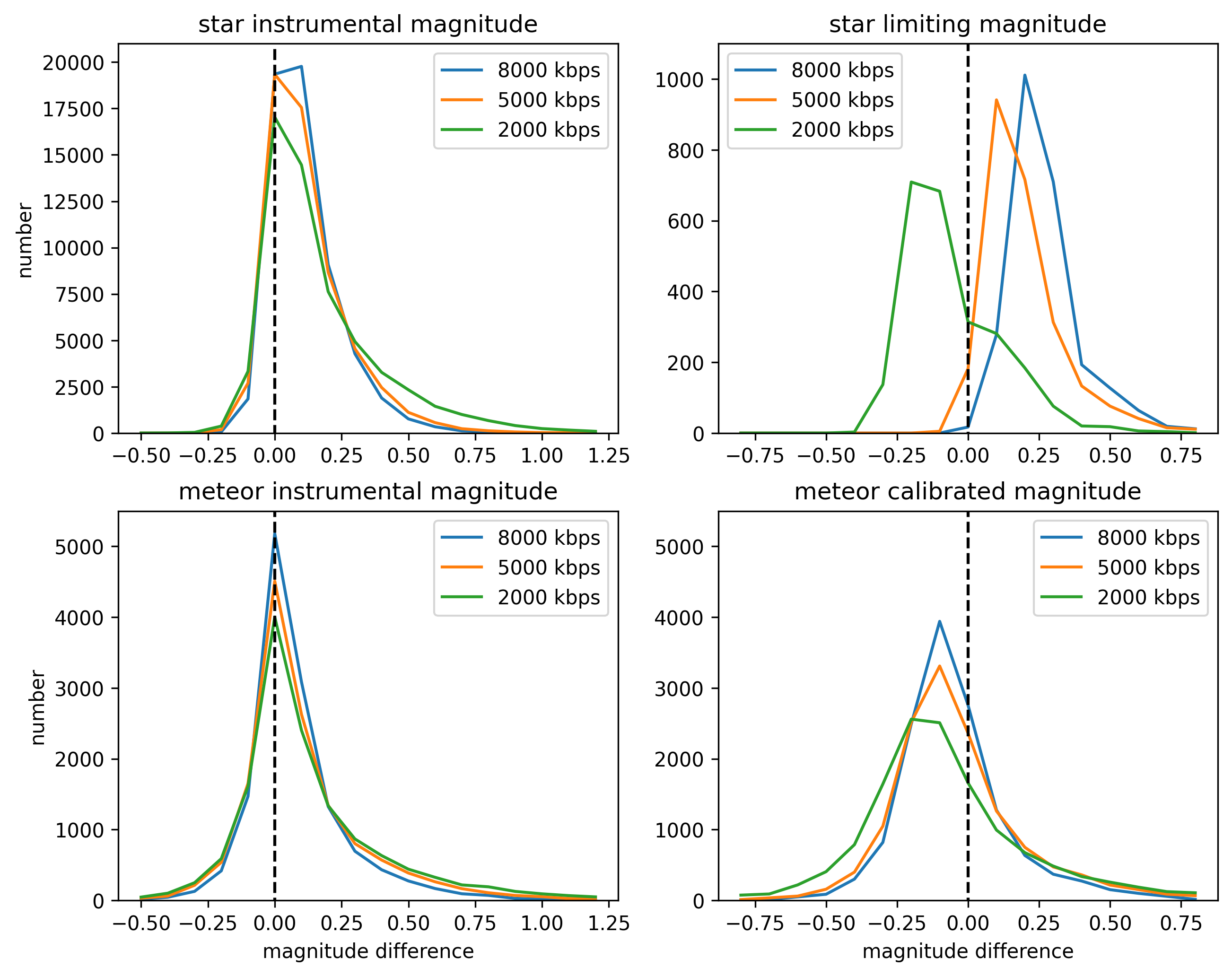}
    \caption{The distribution of the differences of the measured magnitudes between raw and compressed data, which include the instrumental magnitudes of the stars and the meteors, the stellar limiting magnitudes, and calibrated meteor magnitudes.}
    \label{fig:mag_compare}
\end{figure*}

We performed aperture photometry for the stars and the meteors in the raw and compressed data. To demonstrate the difference in the results, the measured magnitudes of the compressed data are subtracted by the ones of the raw data, and the difference of the magnitudes are shown in Figure \ref{fig:mag_compare}. The difference of the instrumental magnitude of the meteors and the stars are shown in the top-left and bottom-left panels of Figure \ref{fig:mag_compare}. We can see that the distribution of the differences are not symmetrical, which indicates that the pixel values are reduced after the compression. However, the reduction of pixel values are not proportional between stars and meteors and between frames, resulting in the deviation of calibrated magnitude of meteors, as shown in the bottom-right panel of Figure \ref{fig:mag_compare}.

The limiting magnitudes are estimated by the SNR of the stars. In this process, the limiting magnitudes are defined as the magnitudes at which half of the stars have SNRs greater than 3. The distributions of the differences of the limiting magnitudes after the compression are shown in the top-right panel of Figure \ref{fig:mag_compare}. As the figure shows, the limiting magnitudes increased after the compression at 8,000 kbps and 5,000 kbps bit-rate. This is because of the reduction of the noises and the overestimate of the SNR of the stars. The limiting magnitude of the compression at 2,000 kbps are reduced, mainly because of the degradation of the image quality.

\subsubsection{Astrometry}

\begin{figure*}[!htb]
    \centering
    \includegraphics[scale=0.8]{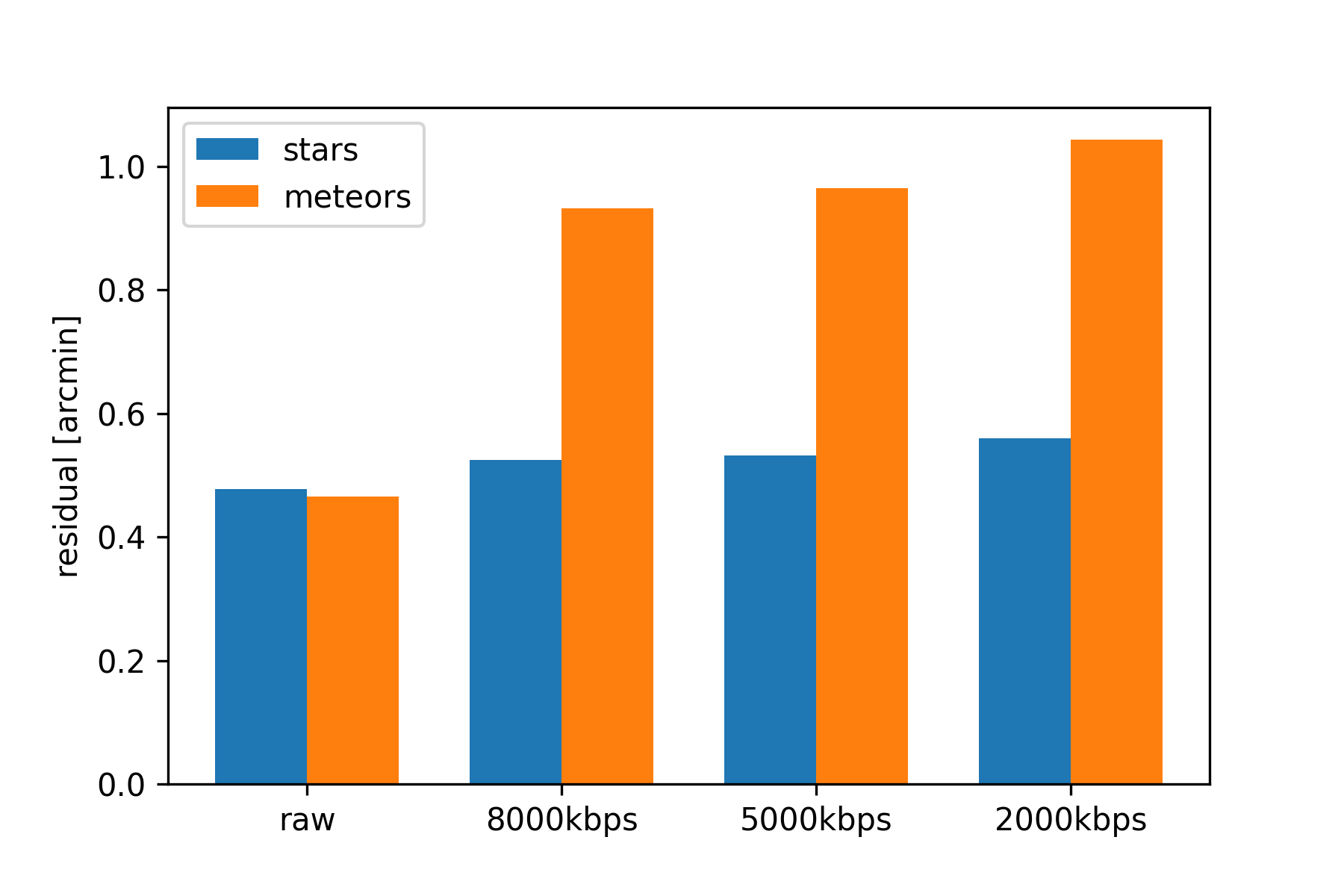}
    \caption{Median of the residuals of stars and meteors by different compression bit-rate. The residuals of stars remain consistent across different bit-rates, and the residuals of meteors are doubled when compression is introduced, and continues to increase as the bit-rate decreases.}
    \label{fig:compare}
\end{figure*}

To test the influence of the astrometric results of the compression, the data structures of meteors with compressed video data are processed by \textit{meteorExtract} and \textit{meteorStitch}, and the results are shown in Figure \ref{fig:compare}. In this figure, we also present the results from the raw data for comparison. From Figure \ref{fig:compare}, we can see that the astrometric residuals of the star is not heavily influenced by the compression, as stars are measured from the averages of dozens of frames, in which defects are smoothed out. However, the fitting residuals of meteoroid trajectories are doubled when compressed, which can be explained by the fact that the meteors are measured separately in each frame when image defects become significant.

\section{Summary} \label{sec:summary}

Meteors are glowing phenomena marked by meteoric entry into the Earth's atmosphere. It is of great importance to the ongoing exploration of the environment around the Earth and interplanetary components in the solar system, as well as research into the origins of interplanetary particles and even the origins of life. Video monitoring is a primary means for meteor observations. It is relatively easy to achieve high accuracy using low-cost hardware. We have designed a new meteor monitoring device, called the Multi-Station Meteor Monitoring system (M$^3$ system), which can be deployed at different sites to form a monitoring network. Also, we have developed a software package that can detect and position the meteors automatically and determine the 3-D trajectories with the data from multiple stations in real time.   

The M$^3$ system consists of a set of lens and camera, a control computer, and a waterproof casing. The set of lens and camera provide a FoV of about 88 $\times$ 58 deg$^2$. The QHY174GPS camera is equipped with a SONY CMOS chip and has a built-in GPS function, which can provide a timing accuracy as high as 1 {\textmu}s. It is a unique feature of our meteor monitoring system, which enables the system to determine the trajectories of meteors accurately. The camera has a frame rate of 30 FPS, with a limiting magnitude of about 4.3 mag. The computer is used to control the camera and process the video data for quick meteor observations and measurements. The elaborate and robust waterproof casing is designed to support the lens-camera set, host the computer, and provide a power supplier. This system can be easily assembled and deployed at sites outside urban areas. 

The matched software package can be used to detect and measure meteors in video streams automatically, aiming to process and save uncompressed data of meteors and other phenomenons. Firstly, we stack the video frames within 2 seconds into photo chunks to identify straight lines, which might be meteors, satellite tracks, or airplanes. The software then classifies the detected lines into different types according to their moving speeds. Secondly, the meteors are positioned in the equatorial coordinate system using the brightest stars in each frame. All the information about meteors will be synchronized to a central server. Finally, the 3-D trajectories and orbits of the meteoroids are determined using the detection information from different sites. 

The M$^3$ systems were tested during the 2021 Geminid meteor shower. We deployed two M$^3$ systems at two sites in the suburbs of Beijing, which are about 55 km apart. We successfully took 4-night observations and detected 700-800 single-station meteors. 473 orbits were computed from these data. The astrometric calibration suggests that the position accuracy is about 0.3-0.4 arcmin. Most meteors are found to be at altitudes of about 80-100 km with an average velocity of about 35.5 ${\rm kms^{-1}}$. According to their radiants and trajectories, 377 of these meteors belong to the Geminid meteoroid stream. The orbits of these meteoroids are quite similar to that of asteroid (3200) Phaethon, which coincides with the knowledge of the origin of this stream. With the data gathered from the test, necessary simulations has been performed to demonstrate the benefit of uncompressed data in the perspective of both astrometry and photometry.

Our meteor systems are highly extensible for building a large-scale monitoring network in China. Some future considerations and possible upgrades are listed below: 
\begin{itemize}
\item \textbf{A lens with a larger aperture.} A 8-mm lens will make the limiting magnitude 1 mag fainter.
\item \textbf{Smaller control computer and waterproof casing.} We try to make the M$^3$ system more compact and lower power consumption. 
\item \textbf{Software upgrade and web platform.} The software will be integrated and upgraded for reliable and automatic operations. A web interface will be designed for visual display of our devices and data products.
\item \textbf{A demonstration network.} We will build a small network around Beijing and carry out special observations to demonstrate the capability of our M$^3$ systems. 
\item \textbf{A national network in China and meteor research.} We will construct a wide-spread network across China to gather a significant volume of meteor data to statistically study the meteor distributions along the Earth's orbit and understand their origins.
\end{itemize}

\subsection*{Disclosures}
The authors have no relevant financial interests in the manuscript and no other potential conflicts of interest to disclose.

\subsection* {Code, Data, and Materials Availability} 

The HIPPARCOS Catalogue is available at \url{https://vizier.cds.unistra.fr/viz-bin/VizieR-3?-source=I/239/hip_main}. The Python modules of Astropy \cite{2022ApJ...935..167A}, Skyfield \cite{2019ascl.soft07024R}, and SciPy \cite{2020SciPy-NMeth} can be installed by ``pip" command in most Python environments. The Python module WMPL \cite{2020MNRAS.491.2688V} can be accessed at \url{https://github.com/wmpg/WesternMeteorPyLib}

Data  for  the  paper  are  freely  available  at  Zenodo repository (\url{https://zenodo.org/records/8379803})\cite{li_zhenye_2023_8379803}. Software (\textit{meteorThread}, \textit{meteorExtract}, \textit{meteorStitch}) developed in this research are still under internal trial, users may contact Zhenye Li (\url{lizhenye@nao.cas.cn}) for access.

\subsection* {Acknowledgments}

We are indebted to the referee for thoughtful comments and
insightful suggestions that improved this paper greatly.

Hu Zou acknowledges the support from the National Key R\&D Program of China (grant No. 2022YFA1602902), National Natural Science Foundation of China (NSFC; grant Nos. 12120101003, 12373010), and Beijing Municipal Natural Science Foundation (grant No. 1222028) and the science research grants from the China Manned Space Project with Nos. CMS-CSST-2021-A02 and CMS-CSST-2021-A04.

Jifeng Liu acknowledges support from the NSFC through grant Nos. of 11988101 and 11933004, and support from the New Cornerstone Science Foundation through the New Cornerstone Investigator Program and the XPLORER PRIZE.


\bibliography{cite}   
\bibliographystyle{spiejour}   


\vspace{2ex}\noindent\textbf{Zhenye Li} is a PhD candidate at the National astronomical observatories, Chinese academy of sciences. He received his BS degree in physics at Beijing Normal University in 2017. His interests include the optical observations of meteors and other transient phenomena, the retrieval of meteorites, and statistical analysis of meteoroids.

\vspace{1ex}
\noindent Biographies and photographs of the other authors are not available.

\listoffigures
\listoftables

\end{spacing}
\end{document}